# Recherche de relations spatio-temporelles : une méthode basée sur l'analyse de corpus textuels

Van Tien NGUYEN, Mauro GAIO, et Christian SALLABERRY

Laboratoire LIUPPA, université de Pau et des pays de l'Adour
Avenue de l'université, B.P.1155, 64013 Pau cedex, France
`{vantien.nguyen, mauro.gaio, christian.sallaberry}@univ-pau.fr`

**Résumé** : Les travaux que nous présentons dans cet article sont réalisés dans le cadre du projet GEONTO. Nous proposons une méthode pour l'enrichissement d'une ontologie géographique à partir de l'analyse automatique d'un corpus textuel composé de récits de voyage. Il s'agit d'une méthode basée sur une approche lexico-syntaxique. En analysant le corpus de texte, nous identifions et détectons des relations n-aires. À partir de ces relations n-aires, nous cherchons les relations spatio-temporelles dans le contexte de descriptions d'itinéraires.

**Mots-clés** : Ingénierie des connaissances, ontologie géographique, traitement automatique de langue, relation n-aire, identification et extraction de relations-naires, web sémantique, relations spatio-temporelles.

## 1  Introduction

Dans le cadre du projet GEONTO[1] nous nous intéressons à la construction et à l'enrichissement des ontologies géographiques à partir de textes provenant d'un corpus constitué de récits de voyage mis à disposition par la médiathèque de Pau.

Grâce à une approche lexico-syntaxique par patrons (Séguéla P. & Aussenac-gilles N., 2000 ; Khelif K. et al, 2006), nous avons mis au point une chaîne de traitement automatique afin de marquer des relations n-aires. Après marquage de toutes les relations n-aires, notre objectif est de déterminer et d'extraire celles qui colportent une sémantique géographique. Dans le cadre de cet article nous nous intéressons aux relations spatio-temporelles dans un contexte de descriptions d'itinéraires.

L'article est organisé comme suit. La partie 2 présente les relations spatio-temporelles dans le contexte de description textuelle des itinéraires. Les relation n-aires et notre méthode pour les extraire sont abordées dans la partie 3. Dans les parties 4 et 5, nous présentons notre hypothèse pour la recherche de relations spatio-temporelles à partir de relation n-aires extraites. Enfin, la partie 6 conclut et propose des perspectives.

---

## 2 Relations spatio-temporelles dans le contexte de description textuelle des itinéraires

Selon (Loustau P. et el, 2008), une entité géographique dans un itinéraire est composée d'une composante spatiale, d'une composante temporelle, et d'une composante thématique ou phénomène.

Les auteurs en s'appuyant sur un modèle d'entités spatiales ont défini cinq relations spatiales principales : la relation métrique (e.g à 10 km de Pau), la relation d'orientation (e.g à l'ouest du Pic de la Fourcanade), la relation de type figure géométrique (e.g dans un triangle Pau, Bordeaux, Toulouse), la relation d'adjacence et l'inclusion (les relations topologiques, e.g près de Pau, au centre de Laruns).

Un modèle équivalent pour les entités temporelles a été proposé par (Le Parc-Lacayrelle A. et al., 2007) avec trois relations principales : la relation d'adjacence, (e.g aux alentours du 10 juillet 1990), la relation d'inclusion (e.g au milieu des années 60), et la relation de distance (e.g 20 ans après le début du siècle).

Dans le modèle d'itinéraire proposé par (Loustau P. et al., 2008) il a montré l'importance des Entités Spatiales (ES) et des verbes de déplacement dans l'évocation du déplacement des acteurs lors du récit de leur voyage. Ce modèle utilise le critère de polarité aspectuelle des verbes introduits par (Boons, 1987) et repris par (Laur, 1991). Il modélise donc les déplacements dans la langue en prenant en compte les verbes de déplacement obligatoirement associés à un acteur, des ES, et une Entité Temporelle (ET). De manière très synthétique, un verbe de déplacement est spécialisé en verbe initial, médian ou final. Dans la langue, il est associé à un Acteur et à une ES au moins (qu'elle soit d'origine, intermédiaire ou de destination). Il est également associé à une ET, ce qui permet d'horodater le déplacement. Par exemple, considérons la phrase "je suis sorti de Pau vers Laruns depuis 3 jours". Le verbe "sortir" est un verbe de déplacement de type initial. De plus, ce verbe est suivi par deux entités spatiales Pau et Laruns. Donc, cela évoque la context d'itinéraire.

En prenant dans un premier temps comme granularité la phrase, nous nous intéressons à identifier les relations spatio-temporelles impliquées dans les itinéraires en nous basant sur les relations n-aires telles que définies par le W3C.

## 3 Relation n-aire et extraction des relation n-aires

### 3.1 Définition

Selon le W3C (Noy et Rector, 2006), une relation n-aire est une relation qui comporte plus de deux arguments. Ces relations peuvent être catégorisées selon quatre cas d'utilisation :
  1. Le *UseCase1* est défini comme étant une extension de la relation binaire entre le sujet et l'objet de la proposition étudiée. La caractéristique de cette

   extension étant l'apport d'informations supplémentaires à la relation. Ces apports pouvant porter sur : la probabilité, la raison, la conséquence, le moyen, etc. Par exemple, la phrase «Nous visitons Pau parce que notre ami y habite.» comporte une relation n-aire entre «nous», la ville de Pau et la raison de notre visite.
2. Le *UseCase2* est également une extension de la relation binaire entre le sujet et l'objet, mais dans ce deuxième cas les informations supplémentaires détaillent l'objet. Par exemple, dans la phrase «Je retrouve le chemin que j'avais suivi pour faire l'ascension du Mont-Perdu.», il y a une relation n-aire entre «Je», le chemin retrouvé, et les informations qui décrivent de manière plus détaillées ce chemin.
3. Dans le *UseCase3*, la relation n-aire lie plusieurs arguments mais aucuns ne peut être considéré comme prioritaire dans la relation. Par exemple, dans la phrase «Le frère de mon ami a quitté Pau, pour une ville près de Lyon, depuis deux semaines.», nous avons une relation n-aire dite «départ», par exemple, avec des arguments tels que la personne impliquée dans le départ (Le frère de mon ami), la ville de départ (Pau), la destination (une ville près de Lyon), et la situation temporelle dans laquelle le départ se réalise (depuis deux semaines).
4. La relation de type *UseCase4* est une relation avec une liste d'arguments qui a pour but de modifier l'objet de la phrase qui contient cette relation. Notons que ce qui est important dans ce cas les arguments doivent être ordonnés. Considérons la phrase "Nous avons visité les monuments comme la tour Eiffel, le cité de l'espace, et le musée Louvre." Supposons que les monuments sont ordonnés en fonction du temps, on a une relation n-aire dite "visiter_monuments" avec une liste des monuments visités tels que la tour Eiffel, le cité de l'espace, et le musée Louvre.

## 3.2 Méthode proposée pour l'identification et l'extraction des relations n-aires à partir de texte

### 3.2.1 Identification des relations n-aires

Grâce aux résultats fournis par une analyse syntaxique (nous avons utilisé pour ce faire l'analyseur Alpage (Cabrera I., 2008)), nous disposons de la liste des relations syntaxiques au sein d'un couple de mots. En se basant sur cette liste nous pouvons identifier les différentes catégories de relations n-aires en construisant des patrons se basant sur des marques linguistiques.

L'idée principale est que chacune des catégories est caractérisée par un ensemble déterminé de relations syntaxiques. Par conséquent, l'identification se fait en repérant l'existence de cet ensemble. Pour une précision accrue nous proposons une divisions des 4 catégories du W3C en sous cas.

Illustrons notre démarche à l'aide d'un sous cas de la catégorie *UseCase3*. Ce cas est caractérisé par des relations syntaxiques déterminantes entre le verbe principal de la phrase et plusieurs prépositions (pour, depuis, par, dans, etc.). Ces dépendances sont indiquées par des marqueurs linguistiques que nous avons trouvés dans la sorties de l'analyseur Alpage.

Par exemple, nous considérons la phrase « Le frère de mon ami a quitté Pau, pour une ville près de Lyon, depuis deux semaines. ». La sortie d'Alpage de cette phrase nous indique une relation syntaxique entre le verbe « quitter » et la préposition « pour », et l'autre relation entre ce verbe et la préposition « depuis ». Alors, une relation n-aire de type *Usecase3* est identifiée. L'extraction de cette relation est présentée par la suite.

### 3.2.2   Algorithme d'extraction des relations n-aires

L'objectif de l'extraction d'une relation n-aire est de récupérer la valeur des arguments de cette relation. Un argument est un groupe de mots. Une relation n-aire est caractérisée par l'ensemble de ses arguments. En se basant sur le fait que chaque phrase peut être considérée comme un graphe orienté, notre algorithme se compose de trois étapes : la première étant la détermination du token pivot, la deuxième étant la construction du graphe correspondant à la phrase, la dernière étant la recherche d'argument sur ce graphe.

Par exemple, la phrase « Le frère de mon ami a quitté Pau, pour une ville près de Lyon, depuis deux semaines.» comporte des relations n-aires. Les tokens pivots sont « frère », « quitté », « Pau », « pour », «ville», « depuis », « semaine ». En appliquant notre algorithme, nous déterminons successivement les arguments de la relation n-aire (Fig. 1).

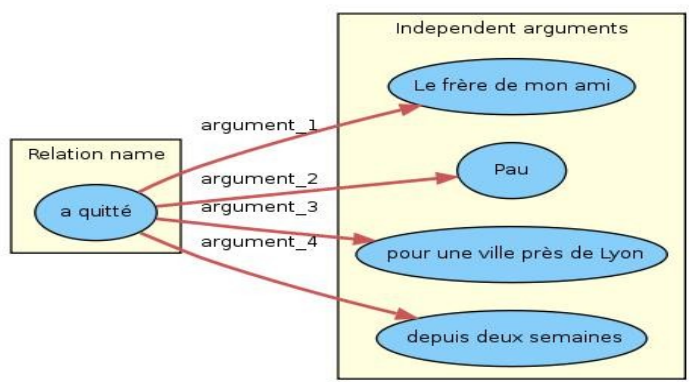

**Fig. 1** -  Un exemple de la relation n-aire extraite

## 4   Les relations n-aires au service des relations spatio-temporelles

Considérons toujours la phrase « Le frère de mon ami a quitté Pau, pour une ville près de Lyon, depuis deux semaines.». Nous avons vu que dans cette phrase, il y a une relation n-aire. L'argument «une ville près de Lyon» de cette relation n-aire comporte une relation spatiale que nous avons définie comme de type *adjacence* que nous avons abordée dans la partie 2. De plus, elle comporte aussi une relation de type de distance temporelle « depuis deux semaines ». Autrement dit, la relation n-aire nous permet de regrouper les entités spatiales et les entités temporelles dans un nouveau lien. Dans ce cas, la relation n-aire nous permet de représenter une relation dite « quitter » entre « Le frère de mon ami », « Pau » (ES), « une ville près de Lyon» (ES), et « depuis deux semaines » (ET).

## 5   Recherche de relation spatio-temporelles à partir des relations n-aires extraites

Dans le corpus de texte, nous traitons l'ensemble des phrases qui comportent une relation n-aire. Pour chacune d'entre elles nous vérifions l'existence d'un verbe de déplacement. Nous cherchons ensuite les ES et les ET de cette relation n-aire. Nous pouvons alors émettre l'hypothèse qu'il existe une relation spatio-temporelle entre ces entités.

Dans notre exemple précédent, *« quitter »* est un verbe de déplacement. Notons que dans certains cas, le verbe *« quitter »* peut porter un autre sens, comme « Il a quitter sa femme, depuis deux semaines ». Pour traiter cette polysémie, nous nous basons sur l'existence des entités spatiale dans la phrase. Dans la phrase « Le frère de mon ami a quitté Pau, pour une ville près de Lyon, depuis deux semaines.», il y a non seulement le verbe de déplacement « quitter », mais encore des entités spatiales nommées « Pau » et « Lyon ».

## 6   Conclusion et travaux futurs

Dans cet article, nous avons montré la méthode d'extraction des relations spatio-temporelles basée sur l'extraction des relation n-aires en utilisant une analyse syntaxique. Dans notre approche, nous supposons que les phrases à traiter comportent des verbes de placement et des entités spatiales nommées. Ces relations seront utilisées dans le but d'enrichir une ontologie géographique.

# Références